\documentclass[aps,onecolumn,amsmath,amssymb,notitlepage,superscriptaddress]{revtex4-1}
\usepackage{graphicx}
\usepackage{dcolumn}
\usepackage{bm}
\usepackage{upgreek}
\usepackage{siunitx}
\usepackage{colonequals}
\usepackage{multirow}

\usepackage[usenames,dvipsnames]{xcolor}

\usepackage{tikz}
\usepackage[kerning=true]{microtype}

\usepackage{xr-hyper}
\definecolor{darkblue}{rgb}{0,0,0.6}
\definecolor{darkred}{rgb}{0.6,0,0}
\usepackage[colorlinks=true,urlcolor=darkblue, citecolor=darkblue, linkcolor=darkred, hyperfootnotes=false]{hyperref}

\usepackage{mathrsfs}

\newcommand{\dd}{\ensuremath{\mathrm d}}

\newcommand{\dep}[2]{\ensuremath{\frac{\partial #1}{\partial #2}}}

\newcommand{\eqlaw}{\ensuremath{\overset{(\mathrm{law})}{=}}}

\def\D{\mathcal{D}}

\def\e{\mathrm{e}}

\def\rt{\tilde{\rho}}
\def\jt{\tilde{j}}

\def\qt{\tilde{q}}
\def\pt{\tilde{p}}

\def\st{\tilde{\sigma}}
\def\Dt{\tilde{D}}

\newcommand{\moy}[1]{\ensuremath{\left\langle #1 \right\rangle}}

\DeclareMathOperator{\erfc}{erfc}

\begin{document}

\title{Duality relation in single-file diffusion}

\author{Pierre Rizkallah}
\affiliation{Sorbonne Universit\'e, CNRS, Physicochimie des Electrolytes et Nanosyst\`emes Interfaciaux (PHENIX), 4 Place Jussieu, 75005 Paris, France}

\author{Aur\'elien Grabsch}
\affiliation{Sorbonne Universit\'e, CNRS, Laboratoire de Physique Th\'eorique de la Mati\`ere Condens\'ee (LPTMC), 4 Place Jussieu, 75005 Paris, France}

\author{Pierre Illien}
\affiliation{Sorbonne Universit\'e, CNRS, Physicochimie des Electrolytes et Nanosyst\`emes Interfaciaux (PHENIX), 4 Place Jussieu, 75005 Paris, France}

\author{Olivier B\'enichou}
\affiliation{Sorbonne Universit\'e, CNRS, Laboratoire de Physique Th\'eorique de la Mati\`ere Condens\'ee (LPTMC), 4 Place Jussieu, 75005 Paris, France}

\date{\today}

\begin{abstract}
    Single-file transport, which corresponds to the diffusion of particles that cannot overtake each other in narrow channels, is an important topic in out-of-equilibrium statistical physics. Various microscopic models of single-file systems have been considered, such as the simple exclusion process (SEP), which has reached the status of a paradigmatic model. Several different models of single-file diffusion have been shown to be related by a duality relation, which holds either microscopically or only in the hydrodynamic limit of large time and large distances.
    Here we show that, within the framework of fluctuating hydrodynamics, these relations are not specific to these models and that, in the hydrodynamic limit, every single-file system can be mapped onto a dual single-file system, which we characterise. This general duality relation allows to obtain new results for different models, by exploiting the solutions that are available for their dual.
\end{abstract}

\maketitle

\section{Introduction}

The study of the dynamical properties of interacting particle systems in out-of-equilibrium situations has been the focus of many works over the last decades~\cite{Evans:2005a,Chou:2011}. The search for a general theory able to describe these systems relies in part on the study of simple microscopic models. This is especially useful for one dimensional systems, for which exact analytical results can be obtained~\cite{Kipnis:1982,Arratia:1983,Derrida:1998,Derrida:2009,Hegde:2014,Krapivsky:2015a,Imamura:2017,Imamura:2021,Poncet:2021,Grabsch:2022}. These 1D models notably describe the important situation of single-file diffusion, which corresponds to the motion of particles in narrow channels with the constraint that they cannot bypass each other. This geometrical constraint has a strong effect on the motion of the particles, which exhibit a subdiffusive behaviour~\cite{Harris:1965,Levitt:1973,Arratia:1983}. This theoretical prediction has been observed at different scales, from the diffusion of molecules in zeolites~\cite{Hahn:1996} to the motion of colloids in narrow trenches~\cite{Wei:2000,Lin:2005}.

\begin{figure}
    \centering
    \includegraphics[width=0.6\textwidth]{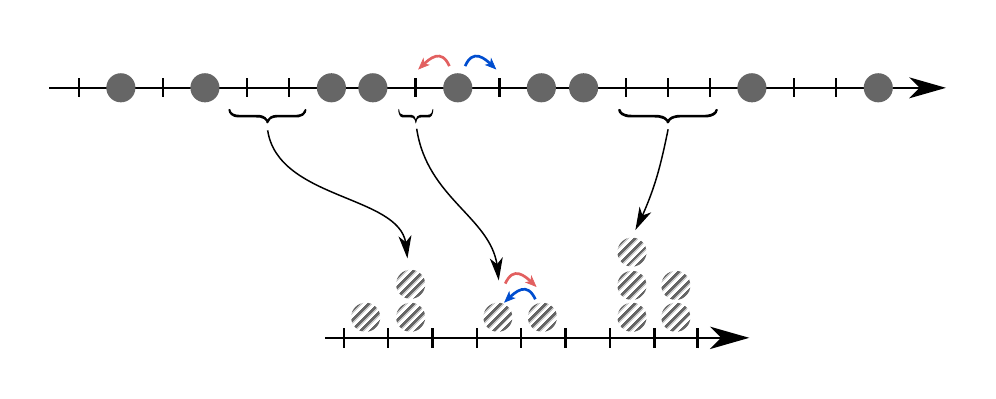}
    \caption{Mapping of the simple exclusion process (SEP, top) onto the zero range process (ZRP, bottom)~\cite{Spitzer:1970,Evans:2000,Evans:2005}. The particles of the SEP can hop on neighbouring sites, only if they are empty. The sites of the ZRP correspond to the particles of the SEP. The number of particles on a site in the ZRP is the number of empty sites on the right of the corresponding particle in the SEP. For the SEP, we have represented by two arrows the possible jumps of an individual particle. We have represented in the ZRP the corresponding dynamics, illustrated by the two arrows corresponding to the jumps of the SEP particle.}
    \label{fig:MappingSEPtoZRP}
\end{figure}

For diffusive systems (satisfying a Fourier law) of particles, the simple exclusion process (SEP) has reached the status of a paradigmatic model~\cite{Mallick:2015}. It consists of particles on a lattice, which can hop to neighbouring sites, with an exclusion rule that enforces that each site can host at most one particle (see Fig.~\ref{fig:MappingSEPtoZRP}, top). The SEP on the infinite line can be solved by Bethe ansatz, and has yielded a number of exact results on the current fluctuation or the displacement of a tagged particle~\cite{Derrida:2009,Derrida:2009a,Imamura:2017,Imamura:2021}.

The SEP is well known to be related to another model of particles on a lattice: the zero range process (ZRP)~\cite{Spitzer:1970} (see the reviews~\cite{Evans:2000,Evans:2005}). In the ZRP, the number of particles on a lattice site is not limited (see Fig.~\ref{fig:MappingSEPtoZRP}, bottom). The rate at which a particle hops from a site can depend on the site and on the number of particles on it, but not on the neighbouring sites. The interaction between the particles thus occur only on the same site (through the hopping rate), hence the name zero range. The SEP can be mapped onto the ZRP with constant hopping rates, by setting the occupations of the sites in the ZRP to be equal to the number of empty sites between the particles of the SEP (see Fig.~\ref{fig:MappingSEPtoZRP}). This relation between the two models has been used explicitly in several works, see for instance~\cite{Landim:1998,Schonherr:2004,Cividini:2017,Lobaskin:2020}.

A similar relation has been made between two other well studied models of one dimensional diffusive systems: the random average process (RAP) and the Kipnis Marchioro Presutti (KMP) model. The RAP is a model of particles on the continuous line, on which particles can hop, in either direction, to a random fraction of the distance to the next particle~\cite{Ferrari:1998}. This model can be mapped onto a mass transfer model~\cite{Ferrari:1998,Krug:2000,Cividini:2016a,Cividini:2016}, which, in the hydrodynamic limit of large time and large distances, becomes equivalent to the KMP model~\cite{Kundu:2016}. The KMP model has been introduced to study the heat conductivity~\cite{Kipnis:1982}. It is again a lattice model, but now each site can host a continuous variable which represents a mass (or an energy), and can randomly exchange mass with its neighbor.

In this article, we show that these relations between different models, which hold either microscopically (for the SEP/ZRP) or in the hydrodynamic limit (for the RAP/KMP), are not specific to the examples mentioned above: by applying these transformations, any single-file system can, in the hydrodynamic limit, be mapped onto another single-file system, which we characterise. This allows to obtain new results on several models of single-file systems by exploiting the solutions available on their dual. For this, we rely on the framework of fluctuating hydrodynamics and macroscopic fluctuation theory (MFT), which is a powerful tool to study the large scale behavior of diffusive systems~\cite{Bertini:2006,Bertini:2007,Bertini:2015}. Its major feature is to allow a unified description of a wide range of systems in terms of only two transport coefficients: a diffusion coefficient $D(\rho)$, and a mobility coefficient $\sigma(\rho)$, which both depend on the local density $\rho$ of the system.
Macroscopic fluctuation theory has been applied with success to the computation of the large deviations of the steady-state current across a diffusive system maintained out-of-equilibrium by two reservoirs~\cite{Bodineau:2004,Bertini:2005a,Bodineau:2005}, and later to study the current on a infinite line with step initial conditions~\cite{Derrida:2009a,Krapivsky:2012}, or to tracer diffusion~\cite{Krapivsky:2015a}. Recently, the MFT equations have been solved exactly in several cases~\cite{Bettelheim:2022,Mallick:2022,Krajenbrink:2022}. 

Note that many microscopic models discussed here have asymmetric versions, such as the asymmetric simple exclusion process (ASEP) in which all the particles have different jump rates to the left and to the right. The mappings onto other models are still possible, but not the description in terms of fluctuating hydrodynamics. Here, we will only focus on the symmetric case, for which the density has a diffusive scaling and fluctuating hydrodynamics can be applied.

\subsection{Summary of the main results}

We show that a single-file system described at large scale by the two coefficients $D(\rho)$ and $\sigma(\rho)$, whose definitions are recalled in Section~\ref{sec:HydroFluc}, can be mapped onto a dual single-file system associated with the transport coefficients
\begin{equation}
    \label{eq:DsigIntro}
    \Dt(\rho) = \frac{1}{\rho^2} D\left( \frac{1}{\rho} \right)
    \:,
    \quad 
    \st(\rho) = \rho \: \sigma \left( \frac{1}{\rho} \right)
    \:.
\end{equation}
The corresponding transformation for the density field $\rho(x,t)$ and the current $j(x,t)$ are written in Section~\ref{sec:SummaryDual}. While these transformations have already been introduced for the SEP/ZRP~\cite{Evans:2000,Evans:2005} and for the RAP/KMP~\cite{Cividini:2016,Kundu:2016}, we stress that our main result is that (i) the transformed density still describes a single-file system, and (ii) this single-file system is explicitly characterised in terms of the original system through~\eqref{eq:DsigIntro}.

We call such a mapping between two single-file systems a duality relation because, as we will argue in Section~\ref{sec:Duality}, the two dual models actually correspond to two equivalent descriptions of the same system. This duality relation is different from the stochastic duality studied in the context of interacting particle systems, such as in~\cite{Carinci:2013}.

We also consider in Section~\ref{sec:TrDil} three other transformations (corresponding to translation or dilatation of the density and current), which, combined with the duality relation described above, allow to map several different single-file systems onto each other. We argue in Appendix~\ref{sec:AppOnlyMappings} that the combination of these four elementary transformations gives all the possible mappings between two single-file systems.

\begin{table}
    \[
\renewcommand{\arraystretch}{2}
\begin{array}{l*2{|>{\displaystyle}c}}
\text{Model} & D(\rho)  & \sigma(\rho)\\ \hline
\text{Simple exclusion process~\cite{Bodineau:2004,Derrida:2009a}} & D_0 & 2 D_0 \rho(1-\rho)  \\
\text{Zero range processes~\cite{Bodineau:2004}} & \frac{\sigma'(\rho)}{2} & \sigma(\rho)\\
\text{Random average process~\cite{Krug:2000,Kundu:2016}} & \frac{\mu_1}{2 \rho^2} & \frac{1}{\rho} \frac{ \mu_1 \mu_2}{\mu_1 - \mu_2}\\
\text{Kipnis-Marchioro-Presutti~\cite{Zarfaty:2016}} & D_0 & \sigma_0 \rho^2\\
\text{Hard Brownian particles~\cite{Krapivsky:2015}} & D_0 & 2D_0 \rho \\
\text{Hard rod gas~\cite{Lin:2005}} & \frac{D_0}{(1-\ell \rho)^2} & 2 D_0 \rho\\
\text{Double exclusion process~\cite{Hager:2001,Baek:2017}} & \frac{D_0}{(1-\rho)^2} & \frac{2 D_0 \rho(1-2\rho)}{1-\rho}
\end{array}
\]
    \caption{Transport coefficients $D(\rho)$ and $\sigma(\rho)$ for the different models studied in this paper (see Section~\ref{sec:Applications} or the caption of Fig.~\ref{fig:Models} for the definitions). For the zero range process we will consider, $\sigma(\rho) = 2 D_0 \rho/(1+\rho)$. For the double exclusion process, the transport coefficients are obtained by taking a limit of the more general Katz–Lebowitz–Spohn (KLS) model, whose transport coefficients are recalled in the Supplementary Material of Ref.~\cite{Baek:2017}.}
    \label{tab:TransportCoefs}
\end{table}

As an application of these transformations, we show that we can obtain new results on several models of single-file diffusion by mapping them onto a model for which a solution has been previously derived. We thereby characterise the statistical properties of two important observables in single-file diffusion (position of a tracer and integrated current through the origin) for several models using the known results that are available on the SEP~\cite{Derrida:2009,Imamura:2017,Imamura:2021,Grabsch:2022}. Indeed, we compute in Section~\ref{sec:Applications} the cumulant generating function of these observables, for the different single-file systems listed in Table~\ref{tab:TransportCoefs}, and represented in Fig.~\ref{fig:Models} below.

\section{Hydrodynamic description}
\label{sec:HydroFluc}

Consider a system of particles, labelled by an index $k$, located at positions $x_k(t)$. These positions can either be continuous variables (e.g. for Brownian particles) or discrete variables (e.g. for the SEP). The time evolution of these positions is stochastic, and depends on the details of the microscopic model. The set of positions $\{ x_k(t) \}$ can be described by a density
\begin{equation}
    \rho(x,t) = \sum_k \delta(x - x_k(t))
    \:.
\end{equation}
At the macroscopic level (large time, large distances), this density can be replaced by a continuous field, which satisfy a continuity relation
\begin{equation}
\label{eq:ContRel}
    \partial_t \rho(x,t) + \partial_x j(x,t) = 0
    \:,
\end{equation}
where $j(x,t)$ is the local current at position $x$ and time $t$. The stochastic behaviour of the underlying microscopic model can be taken into account by assuming that the current $j$ itself is random, and takes the form~\cite{Spohn:1991}
\begin{equation}
\label{eq:StochCurrent}
    j(x,t) = - D(\rho)\partial_x \rho(x,t) + \sqrt{\sigma(\rho)} \: \eta(x,t)
    \:,
\end{equation}
where $\eta$ is a Gaussian white noise in space and time with unit variance, i.e.,
\begin{equation}
    \moy{\eta(x,t) \eta(x',t')} = \delta(x-x') \delta(t-t')
    \:.
\end{equation}
This strength of this approach, named fluctuating hydrodynamics, is that all the microscopic details of the model (displacement, interaction, ...) are encoded into two transport coefficients:
\begin{itemize}
    \item the diffusion coefficient $D(\rho)$;
    \item the mobility $\sigma(\rho)$, which characterises the amplitude of the fluctuations of the current $j$.
\end{itemize}
These coefficients can be computed by considering a finite system of length $L$, placed between two reservoirs at density $\rho_1$ and $\rho_2$~\cite{Derrida:2011}. $D(\rho)$ and $\sigma(\rho)$ are related to the first two moments of the total current $Q_t$ that flows between the reservoirs:
\begin{equation}
    D(\rho) = \lim_{L \to \infty}
    \lim_{\rho_1 \to \rho_2 = \rho} \frac{L}{\rho_1-\rho_2} \lim_{t\to \infty} \frac{\moy{Q_t}}{t}
    \:,
    \quad
    \sigma(\rho) = \lim_{L \to \infty}
     \lim_{\rho_1 \to \rho_2 = \rho} L \lim_{t\to \infty} \frac{\moy{Q_t^2}}{t}
    \:.
\end{equation}
In practice, these two coefficients can be difficult to compute analytically. Nevertheless, they have been obtained for different microscopic models (see for instance Table~\ref{tab:TransportCoefs}).

\section{Particles/gaps duality relation}
\label{sec:Duality}

We now show that there is a duality relation between different systems, at the level of the fluctuating hydrodynamics. This duality relation is general, and can be applied to any single-file system (not necessarily systems of particles). Nevertheless, we first define the dual fields $\rt$ and $\jt$, from the original fields $\rho$ and $j$, by using heuristic arguments valid for systems of particles (section~\ref{sec:DefDualFields}). These definitions coincide with the ones used in Refs.~\cite{Evans:2000,Evans:2005} for the SEP/ZRP, and in~\cite{Cividini:2016,Kundu:2016} for the RAP/KMP. We then proceed to show that, for any single-file system, with these definitions, the dual fields \textit{still} obey the equations of fluctuating hydrodynamics (section~\ref{sec:MappingDual}), with new transport coefficients, which we determine.

\subsection{Dual density and current fields}
\label{sec:DefDualFields}

\begin{figure}
	\centering
	\includegraphics[width=0.8\textwidth]{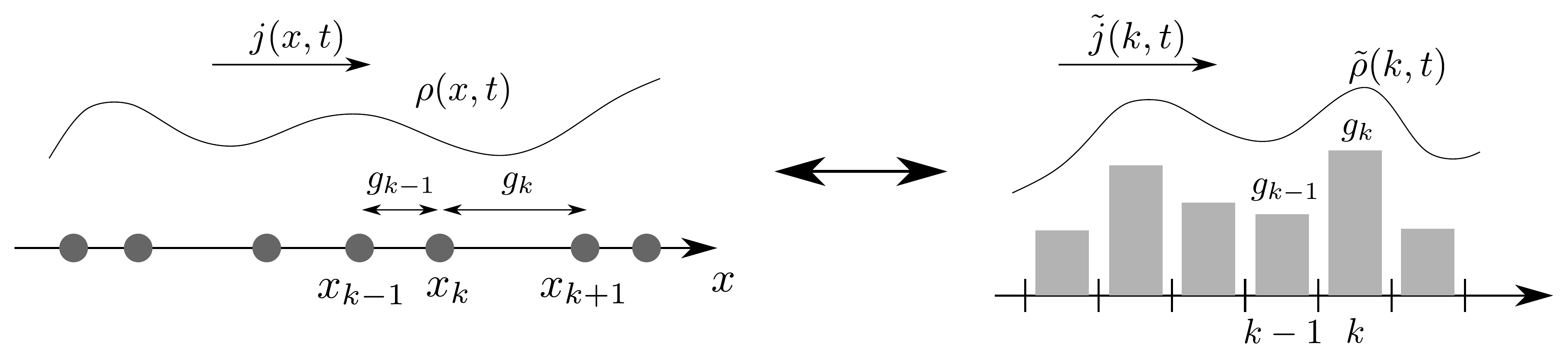}
	\caption{Duality relation between two equivalent descriptions of a particle system: through the positions of the particles (left) and the through the gaps between the particles (right). The particle system is described at the macroscopic level by a density field $\rho(x,t)$ and a current $j(x,t)$, while the dual system (gaps) is described by a density $\rt(k,t)$ and a current $\jt(k,t)$. These two descriptions are equivalent, and related by Eqs.~(\ref{eq:DualFields},\ref{eq:DualPos}). The inverse transformation is identical, and written explicitly in Eqs.~(\ref{eq:DualToFirst},\ref{eq:PosFromDual}). These two sets of fields $(\rho,j)$ and $(\rt,\jt)$ obey the equations of fluctuating hydrodynamics~(\ref{eq:ContRel},\ref{eq:StochCurrent}), with transport coefficients respectively given by $(D,\sigma)$ and $(\Dt,\st)$, which are related by the transformation~\eqref{eq:RelCoefsTransp}.}
	\label{fig:mapping}
\end{figure}

Let us consider a system of particles at positions $\{ x_k(t) \}$, described at the macroscopic level by the two continuous fields $\rho(x,t)$ and $j(x,t)$, and the two transport coefficients $D(\rho)$ and $\sigma(\rho)$. The idea behind the duality relation is that this system of particles can be equivalently described by considering the gaps $\{ g_k = x_{k+1} - x_k \}$ between the particles, as shown in Fig.~\ref{fig:mapping}. Note that we will introduce the new fields only at the macroscopic level, at which $k$ can be considered as a continuous variable. We will however motivate the definitions of these fields by microscopic considerations (in which $k$ represents the label of a particle and is in principle an integer). At large scale, the gaps are described by a density field
\begin{equation}
    \label{eq:ContGaps}
    \rt(k,t) = g_k \equiv x_{k+1}(t) - x_{k}(t)
\end{equation}
and the associated current $\jt(k,t)$. Microscopically, $k$ is a discrete variable representing the label of the particle. Here, at the macroscopic level, it is a continuous variable which interpolates between these discrete values. The system of particles obeys the equations of the fluctuating hydrodynamics~(\ref{eq:ContRel},\ref{eq:StochCurrent}):
\begin{align}
\label{eq:ContRel2}
    \partial_t \rho(x,t) + \partial_x j(x,t) &= 0
    \:,
    \\
\label{eq:defjfluc2}
    j(x,t) + D(\rho)\partial_x \rho(x,t) & = \sqrt{\sigma(\rho)} \: \eta(x,t)
    \:.
\end{align}
This is our starting point to obtain the equations satisfied by $\rt$ and $\jt$.

\begin{figure}
    \centering
    \includegraphics[width=0.8\textwidth]{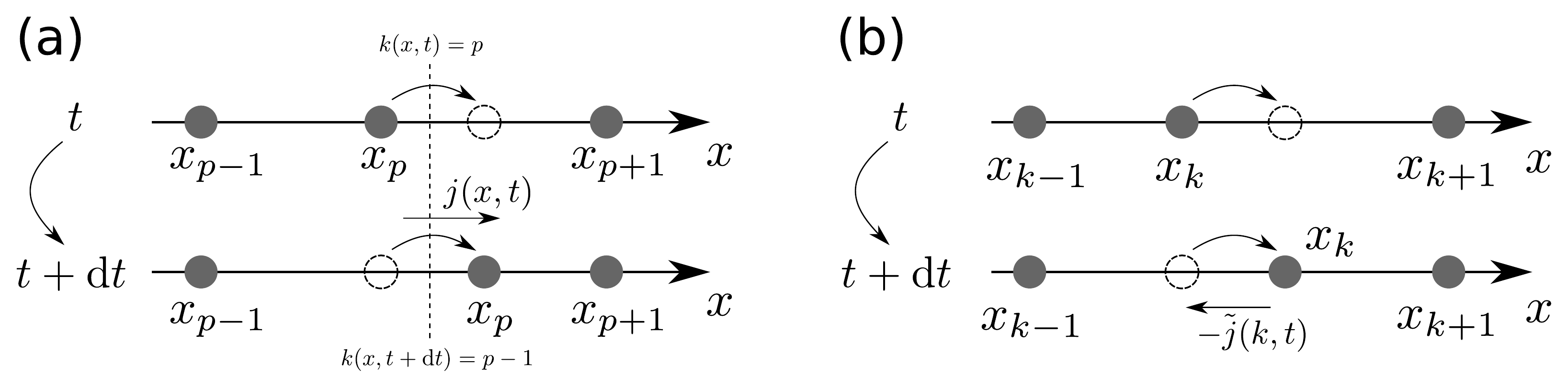}
    \caption{\textbf{(a)} the flux $j(x,t)$ of particles at position $x$ at time $t$ gives the variation of the index $k(x,t)$ of the closest particle to the left of $x$, see Eq.~\eqref{eq:Evolkt}. \textbf{(b)} The flux $\jt(k,t)$ of gaps through particle $k$ at time $t$ gives the evolution of the position of this particle~\eqref{eq:defjtviaxt}.}
    \label{fig:FluxGaps}
\end{figure}

The first step is to express the label $k(x,t)$ of the particle located at position $x$ at time $t$ in terms of the density $\rho$. Microscopically, this is a piecewise constant function, which gives the index of the closest particle to the left of $x$ at time $t$. At the macroscopic level, this becomes a continuous function, which can be defined by writing the number of particles between $0$ and $x$ as
\begin{equation}
\label{eq:indexkt}
    k(x,t) - k(0,t) = \int_0^x \rho(x',t) \dd x'
    \:.
\end{equation}
That requires the knowledge of $k(0,t)$, which is the index of the particle located at $x=0$. It can be determined by writing that the variation of $k(x,t)$ during a short time $\dd t$ corresponds to the number of particles that crossed $x$ from right to left, i.e. $-j(x,t) \dd t$ (see Fig.~\ref{fig:FluxGaps}(a)). Therefore,
\begin{equation}
\label{eq:Evolkt}
    \partial_t k(x,t) = -j(x,t)
    \:.
\end{equation}
Note that this is coherent with~\eqref{eq:indexkt} and the continuity relation~\eqref{eq:ContRel2}. Taking the convention that at $t=0$, the particle located at $x=0$ has index $k(0,0) = 0$, Eqs.~(\ref{eq:indexkt},\ref{eq:Evolkt}) fully determine $k(x,t)$ for all $x$ and $t$.

We now relate the density fields $\rho$ and $\rt$ by defining the density of particles as the inverse of the gap between the particles located at position $x$ at time $t$,
\begin{equation}
\label{eq:gapsdensity}
    \rho(x,t) = \frac{1}{g_{k(x,t)}} = \frac{1}{\rt(k(x,t),t)}
    \:,
\end{equation}
from Eq.~\eqref{eq:ContGaps}. The current $\jt(k,t)$ corresponds to the "flux of gap" through the particle $k$, which is thus given by the motion of this particle (see Fig.~\ref{fig:FluxGaps}(b)),
\begin{equation}
\label{eq:defjtviaxt}
    \jt(k,t) = -\partial_t x_k(t)
    \:.
\end{equation}
We can express $\partial_t x_k$ in terms of $\rho$ and $j$ by writing that the number of particles between $x_p(t)$ and $x$ is
\begin{equation}
    \int_{x_p(t)}^{x} \rho(x',t) \dd x' = k(x,t) - k(x_p(t),t) = k(x,t) - p
    \:,
\end{equation}
since $k(x_p(t),t) = p$. Taking the derivative of this relation with respect to $t$, and using Eqs.~(\ref{eq:ContRel2},\ref{eq:Evolkt}) we obtain
\begin{equation}
    j(x_p(t),t) -  \rho(x_p(t),t) \: \partial_t x_p(t) = 0
    \:.
\end{equation}
Combining this relation with~\eqref{eq:defjtviaxt} yields
\begin{equation}
\label{eq:Reljtrhoj}
    \jt(k(x,t),t) = - \frac{j(x,t)}{\rho(x,t)}
    \:.
\end{equation}

Equations~(\ref{eq:indexkt},\ref{eq:Evolkt}) define the index $k(x,t)$, while Eqs.~(\ref{eq:gapsdensity},\ref{eq:Reljtrhoj}) define the new fields $\rt(k,t)$ and $\jt(k,t)$ from the original fields $\rho(x,t)$ and $j(x,t)$. The same definitions can be found in Refs.~\cite{Evans:2000,Evans:2005} for the SEP/ZRP, and in~\cite{Cividini:2016,Kundu:2016} for the RAP/KMP. Although here we have introduced these definitions based on physical arguments for a system of particles, the dual fields can be defined from any fields $\rho(x,t)$ and $j(x,t)$ which obey the equations of fluctuating hydrodynamics~(\ref{eq:ContRel2},\ref{eq:defjfluc2}).
We will now show that these new fields $\rt(k,t)$ and $\jt(k,t)$ still obey these equations, with new transport coefficients.

\subsection{Mapping the equations of fluctuating hydrodynamics}
\label{sec:MappingDual}

Let us now deduce the new equations satisfied by $\rt$ and $\jt$ from the fluctuating hydrodynamics equations~(\ref{eq:ContRel2},\ref{eq:defjfluc2}). We first rewrite~\eqref{eq:Reljtrhoj} using~\eqref{eq:gapsdensity} as
\begin{equation}
\label{eq:Exprj2}
    j(x,t) = -\frac{\jt(k(x,t),t)}{\rt(k(x,t),t)}
    \:.
\end{equation}
Taking the time derivative of Eq.~\eqref{eq:gapsdensity} we get
\begin{align}
    \partial_t \rho(x,t) 
    &= \partial_t \frac{1}{\rt(k,t)} + \partial_t k(x,t) \: \partial_k \frac{1}{\rt(k,t)}
    \nonumber 
    \\
    &= - \frac{1}{\rt(k,t)^2} \left( \partial_t \rt(k,t) - j(x,t) \partial_k \rt(x,t) \right)
    \nonumber 
    \\
    &= - \frac{1}{\rt(k,t)^2} \left( \partial_t \rt(k,t) + \frac{\jt(k,t)}{\rt(k,t)} \partial_k \rt(x,t) \right)
    \label{eq:dtrho}
\end{align}
where we have used~\eqref{eq:Evolkt} and~\eqref{eq:Exprj2}. Similarly,
\begin{equation}
\label{eq:dxrho}
    \partial_x \rho(x,t) = \partial_x k(x,t) \: \partial_k \frac{1}{\rt(k,t)}
    = -\frac{1}{\rt(k,t)^3} \partial_k \rt(k,t)
    \:,
\end{equation}
\begin{equation}
\label{eq:dxj}
    \partial_x j(x,t) = - \frac{1}{\rt(k,t)} \partial_k \frac{\jt(k,t)}{\rt(k,t)}
    = - \frac{1}{\rt(k,t)^2} \left(
        \partial_k \jt(k,t) - \jt(k,t) \frac{\partial_k \rt(k,t)}{\rt(k,t)}
    \right)
    \:.
\end{equation}
Plugging~(\ref{eq:dtrho},\ref{eq:dxj}) into the continuity relation~\eqref{eq:ContRel}, we obtain a new continuity equation for the fields $\rt$ and $\jt$:
\begin{equation}
\label{eq:ContinDual}
    \partial_t \rt(k,t) + \partial_k \jt(k,t) = 0
    \:.
\end{equation}
Similarly, using~\eqref{eq:dxrho} into the expression of the current~\eqref{eq:defjfluc2}, we obtain
\begin{equation}
    \jt(k,t) + D \left(\frac{1}{\rt} \right) \frac{1}{\rt^2(k,t)} \partial_k \rt(k,t)
    = -\rt(k,t) \sqrt{\sigma \left( \frac{1}{\rt} \right)} \: \eta(x_k(t),t)
\end{equation}
The transformation of the Gaussian white noise $\eta(x,t)$ under the change of variables can be obtained from the distribution,
\begin{equation}
    P[\eta] \propto \exp \left[ 
    - \int_{0}^T \dd t \int_{-\infty}^\infty \dd x \: \frac{\eta(x,t)^2}{2}
    \right]
    = \exp \left[ 
    - \int_{0}^T \dd t \int_{-\infty}^\infty \dd k \: \dep{x_k}{k} \frac{\eta(x_k(t),t)^2}{2}
    \right]
    \:.
\end{equation}
Therefore, $\eta(x_k(t),t)$ is a Gaussian white noise with variance $\dep{k}{x}$, which we denote
\begin{equation}
    \eta(x_k(t),t) \eqlaw \sqrt{\dep{k}{x}} \: \eta(k,t)
    = \frac{1}{\sqrt{\rt(k,t)}} \eta(k,t)
    \:,
\end{equation}
where $\eqlaw$ indicates that the two noises have the same distribution, and $\eta(k,t)$ is also a Gaussian white noise with unit variance. We thus get (using $\eta(k,t) \eqlaw - \eta(k,t)$),
\begin{equation}
\label{eq:FlucHydDual}
    \jt(k,t) + \Dt(\rt) \partial_k \rt(k,t) = \sqrt{\st(\rt)} \: \eta(k,t)
    \:,
\end{equation}
where we have denoted
\begin{equation}
\label{eq:RelCoefsTransp0}
    \Dt(\rt) = \frac{1}{\rt^2} D\left( \frac{1}{\rt} \right)
    \:,
    \quad 
    \st(\rt) = \rt \: \sigma \left( \frac{1}{\rt} \right)
    \:.
\end{equation}
The dual system, which describes the dynamics of the gaps, therefore obeys the equations of fluctuating hydrodynamics~(\ref{eq:ContinDual},\ref{eq:FlucHydDual}), with transport coefficients $\Dt$ and $\st$ which are related to the original coefficients $D$ and $\sigma$ of the particle system by the transformations~\eqref{eq:RelCoefsTransp0}. The known duality relation between the RAP and the KMP model, used in Refs.~\cite{Cividini:2016a,Cividini:2016,Kundu:2016} is a specific case of the duality relation described here, applied to the transport coefficients of these models, given in Table~\ref{tab:TransportCoefs}.

\subsection{Summary and discussion}
\label{sec:SummaryDual}

We have started from a system of particles, and showed that the system can be equivalently described in terms of the gaps between the particles. The two systems are related by the transformation
\begin{equation}
\label{eq:DualFields}
    \rho(x,t) = \frac{1}{\rt(k(x,t),t)}
    \:,
    \qquad
    j(x,t) = -\frac{\jt(k(x,t),t)}{\rt(k(x,t),t)}
    \;,
\end{equation}
where
\begin{equation}
\label{eq:DualPos}
    k(x,t) - k(0,t) = \int_0^x \rho(x',t) \dd x'
    \;,
    \qquad
    \partial_t k(x,t) = -j(x,t)
    \:,
\end{equation}
which maps the original equations~(\ref{eq:ContRel},\ref{eq:StochCurrent}) on $\rho$ and $j$, onto the same equations of fluctuating hydrodynamics, but with the new transport coefficients
\begin{equation}
\label{eq:RelCoefsTransp}
    \Dt(\rt) = \frac{1}{\rt^2} D\left( \frac{1}{\rt} \right)
    \:,
    \quad 
    \st(\rt) = \rt \: \sigma \left( \frac{1}{\rt} \right)
    \:.
\end{equation}

Although we have derived these relations by using physical arguments, from a mathematical point of view, they can be applied to any system which obeys the equations~(\ref{eq:ContRel},\ref{eq:StochCurrent}) of fluctuating hydrodynamics, not necessarily systems of particles. In particular, by applying this transformation to the dual system, we recover the original system. Explicitly, this means that the inverse transformation of the current and density is identical:
\begin{equation}
\label{eq:DualToFirst}
    \rt(k,t) = \frac{1}{\rho(x_k(t),t)}
    \:,
    \qquad
    \jt(k,t) = -\frac{j(x_k(t),t)}{\rho(x_k(t),t)}
    \:,
\end{equation}
with
\begin{equation}
\label{eq:PosFromDual}
    x_k(t) - x_0(t) = \int_0^k \tilde{\rho}(k',t)\dd k'
    \:,
    \qquad
    \partial_t x_k(t) = - \jt(k,t)
    \:.
\end{equation}

Note that these relations, which are valid on the stochastic equations of fluctuating hydrodynamics, also imply a similar duality relation relation on the equations of Macroscopic Fluctuation Theory (MFT), which is a deterministic rewriting of the fluctuating hydrodynamics. We give more details on this relation in Appendix~\ref{sec:AppMFT}.

The transformations presented in this article can be applied to both finite and infinite systems. In the case of a finite system, one should carefully study how the transformations act on the boundary conditions.

We will show in Section~\ref{sec:Applications} that this duality relation can be used to obtain results on a single-file system by studying its dual. But first, we discuss other transformations of the density $\rho$ and current $j$ that will be useful in the following.

\section{Translation and dilatations}
\label{sec:TrDil}

We introduce 3 other transformations, acting on the fields $\rho$ and $j$, which, combined with the duality relation above, give all the possible mapping between two single-file systems. This is shown in Appendix~\ref{sec:AppOnlyMappings}.

\begin{itemize}
    \item \textbf{Translation of density (T).} The first transformation consists in shifting the density by a constant $c$. We define new density and current fields as
    \begin{equation}
        \rt(x,t) = \rho(x,t) + c
        \:,
        \qquad
        \jt(x,t) = j(x,t)
        \:.
    \end{equation}
    These two fields still obey the equations of fluctuating hydrodynamics~(\ref{eq:ContRel},\ref{eq:StochCurrent}), upon changing the transport coefficients as
    \begin{equation}
        \Dt(\rt) = D(\rt - c)
        \:,
        \qquad
        \st(\rt) = \sigma(\rt - c)
        \:.
    \end{equation}
    \item \textbf{Dilatation of the fields (Di).} This transformation multiplies the two fields by a constant $c$. We define
        \begin{equation}
        \rt(x,t) = c \: \rho(x,t)
        \:,
        \qquad
        \jt(x,t) = c \: j(x,t)
        \:,
    \end{equation}
    which obey the equations of fluctuating hydrodynamics~(\ref{eq:ContRel},\ref{eq:StochCurrent}), with the transport coefficients
    \begin{equation}
        \Dt(\rt) = D \left( \frac{\rt}{c} \right)
        \:,
        \qquad
        \st(\rt) = c^2 \: \sigma\left( \frac{\rt}{c} \right)
        \:.
    \end{equation}
    \item \textbf{Rescaling of time (Rt).} The last transformation corresponds to changing the time scale by a constant $\tau$. We introduce
    \begin{equation}
        \rt(x,t) = \rho(x,\tau t)
        \:,
        \qquad
        \jt(x,t) = \tau \: j(x, \tau t)
        \:.
    \end{equation}
    Note that $j$ needs to be rescaled by $\tau$ in order to satisfy the continuity equation~\eqref{eq:ContRel}. These new fields again obey the equations of fluctuating hydrodynamics~(\ref{eq:ContRel},\ref{eq:StochCurrent}), with the transport coefficients
    \begin{equation}
        \Dt(\rt) = \tau \: D (\rt)
        \:,
        \qquad
        \st(\rt) = \tau \: \sigma(\rt)
        \:.
    \end{equation}
\end{itemize}

\section{Applications: integrated current and tracer diffusion}
\label{sec:Applications}

We now focus on the case of an infinite system, and consider two different observables, which have been the focus of many studies in the context of single-file diffusion~\cite{Rajesh:2001,Derrida:2009,Derrida:2009a,Sadhu:2015,Krapivsky:2015a,Imamura:2017,Imamura:2021,Poncet:2021,Grabsch:2022,Mallick:2022}:
\begin{itemize}
    \item The integrated current though the origin $Q_t$, which counts the total number of particles that crossed the origin from left to right (minus the number from right to left) up to time $t$,
    \begin{equation}
        \label{eq:DefQt}
        Q_t = \int_0^t j(0,t')\dd t' = \int_0^\infty (\rho(x,t)-\rho(x,0)) \dd x
        \:.
    \end{equation}
    \item The position $x_0(t)$ of a tagged particle (a tracer), which was initially placed at the origin. This position can be deduced from the density $\rho$ only, using that the number of particles located to the right of $x_0(t)$ is conserved~\cite{Krapivsky:2015a}:
    \begin{equation}
    \label{eq:defx0}
        \int_0^{x_0(t)} \rho(x,t) \dd x =
        \int_0^\infty (\rho(x,t)-\rho(x,0)) \dd x
        \:.
    \end{equation}
\end{itemize}
We will consider the cumulant generating functions associated to these two observables, which in the long time limit take the form,
\begin{equation}
\label{def:CGFs}
    \ln \moy{ \e^{\lambda Q_t}} \simeq \sqrt{t} \: \psi_Q(\lambda)
    \:,
    \qquad
    \ln \moy{ \e^{\lambda x_0(t)}} \simeq \sqrt{t} \: \psi_T(\lambda)
    \:.
\end{equation}
The rescaled cumulant generating functions $\psi_Q$ and $\psi_T$ have been computed for different models of single file systems~\cite{Derrida:2009,Derrida:2009a,Sadhu:2015,Imamura:2017,Imamura:2021}. We now show that, using the transformations we have identified above, the cumulants of these two observables for several other single-file systems can be deduced from these results. We first show how these transformations can be used to relate known results on the SEP and the KMP. Then, we apply them to obtain new results on the different single-file systems shown in Fig.~\ref{fig:Models}.

\subsection{Transformation of the observables}

In order to use the different mappings identified in Section~\ref{sec:Duality} and~\ref{sec:TrDil}, we first study how they transform the two observables $x_0(t)$ and $Q_t$.
As in the previous sections, we consider a model described by the density $\rho$ and current $j$, which is mapped to a new single-file system by one of the transformations defined above. The new system is described by a density $\rt$ and a flux $\jt$. We also denote $\tilde{x}_0(t)$ the position of a tracer in the new system, defined as in~\eqref{eq:defx0} with $\rho$ replaced by $\rt$, and the flux through the origin in the new system is denoted $\tilde{Q}_t$, defined as in~\eqref{eq:DefQt}.
\begin{itemize}
    \item \textbf{Duality relation (D).} We can express the position of the tracer in terms of the integrated current of the dual system, using Eq.~\eqref{eq:PosFromDual}:
    \begin{equation}
        \label{eq:x0fromQdual}
        x_0(t) = x_0(0) - \int_0^t \jt(0,t') \dd t' = - \tilde{Q}_t
    \:,
    \end{equation}
    since $x_0(0) = 0$. The position of the tracer is exactly the opposite of the flux through the origin in the dual model. Applying this relation to the dual system, we also obtain
    \begin{equation}
        \label{eq:x0DualFromQ}
        \tilde{x}_0(t) = - Q_t
        \:.
    \end{equation}
    \item \textbf{Translation (T).} This transformation conserves the current $j$, therefore the integrated current is unchanged, $\tilde{Q}_t = Q_t$.
    For the position $x_0(t)$ of the tracer, Eq.~\eqref{eq:defx0} gives
    \begin{equation}
        \int_{0}^{x_0(t)} (\rt(x,t) - c) \dd x
        = \int_{0}^{\infty} (\rt(x,t) - \rt(x,0) ) \dd x
        \equiv
        \int_{0}^{\tilde{x}_0(t)} \rt(x,t) \dd x
        \:.
    \end{equation}
    This yields the relation
    \begin{equation}
    \label{eq:TransXtTransl}
        \int_{x_0(t)}^{\tilde{x}_0(t)} \rt(x,t) \dd x
        = - c \: x_0(t)
        \:.
    \end{equation}
    \item \textbf{Dilatation (Di).} This transformation multiplies both $\rho$ and $j$ by a constant $c$. Therefore, the integrated current is also multiplied by $c$, $\tilde{Q}_t = c Q_t$, while due to~\eqref{eq:defx0}, the position of the tracer is unchanged: $\tilde{x}_0(t) = x_0(t)$.
    \item \textbf{Rescaling of time (Rt).} This transformation only changes the time scale by a factor $\tau$ (and the current accordingly), such that both the integrated current and the position of the tracer are only rescaled in time: $\tilde{Q}_t = Q_{t \tau}$ and $\tilde{x}_0(t) = x_0(t \tau)$.
\end{itemize}

\begin{figure}
	\centering
	\includegraphics[width=0.65\textwidth]{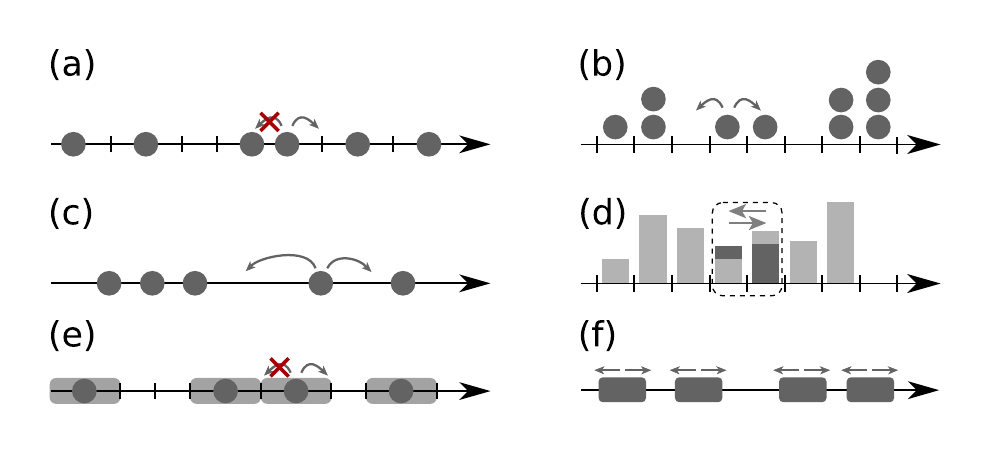}
	\caption{The different models of single-file systems considered in this paper. \textbf{(a)} the simple exclusion process (SEP). Each particle can hop to a neighbouring empty site. \textbf{(b)} the zero range process (ZRP). The particles can hop on a neighbouring site. \textbf{(c)} the random average process (RAP). Particles on a continuous line can hop to a random fraction of the distance to the next particle. \textbf{(d)} the Kipnis Marchioro Presutti model (KMP). Each site hosts a continuous variable (a mass). At random times, the total mass of two neighbouring sites is randomly redistributed between them.  \textbf{(e)} the double exclusion process. A particle can hop on a neighbouring site if the destination site and the next one is empty. It represents an exclusion process with particles that occupy the volume of two sites (shaded region). \textbf{(f)} the gas of hard rods. Particles of length $\ell$ (rods) perform a Brownian motion, with the condition that the rods cannot overlap. The corresponding transport coefficients for these models are given in Table~\ref{tab:TransportCoefs}.}
	\label{fig:Models}
\end{figure}

These transformations are summarised in Table~\ref{tab:TransfOnObs}. We can now apply these transformations to study the different models of single-file diffusion represented in Fig.~\ref{fig:Models}.

\begin{table}
    \[
\renewcommand{\arraystretch}{2}
\begin{array}{l*2{|>{\displaystyle}c}p{0.4cm}l*2{|>{\displaystyle}c}}
\text{Transformation} & \text{Original system}  & \text{Mapped system}
& \multirow{8}{*}{} &
\text{Transformation} & \text{Original system}  & \text{Mapped system}\\
\cline{1-3} \cline{5-7} 
\multirow{5}{*}{\text{Duality relation (Du)}} & D(\rho) & \Dt(\rho) =  \frac{1}{\rho^2} D \left( \frac{1}{\rho} \right)
&&\multirow{5}{*}{\text{Dilatation (Di)}} & D(\rho) & \Dt(\rho) =  D(\rho/c)\\
    & \sigma(\rho) & \st(\rho) =  \rho \: \sigma \left( \frac{1}{\rho} \right)
    && & \sigma(\rho) & \st(\rho) = c^2 \: \sigma(\rho/c)\\
    & \rho & \tilde{\rho} = \frac{1}{\rho}
    &&& \rho & \tilde{\rho} = c \: \rho \\
    & x_0(t) & \tilde{x}_0(t) = - Q_t
    && & x_0(t) & \tilde{x}_0(t) = x_0(t) \\
    & Q_t & \tilde{Q}_t = -x_0(t)
    &&& Q_t & \tilde{Q}_t = c \: Q_t\\
\cline{1-3} \cline{5-7} 
\multirow{4}{*}{\text{Translation (T)}} & D(\rho) & \Dt(\rho) = D(\rho-c)
&& \multirow{5}{*}{\text{Rescaling time (Rt)}} & D(\rho) & \Dt(\rho) = \tau \: D(\rho)\\
    & \sigma(\rho) & \st(\rho) = \sigma(\rho-c)
    && & \sigma(\rho) &  \st(\rho) = \tau \: \sigma(\rho)\\
    & \rho & \tilde{\rho} = \rho + c
    &&& \rho & \tilde{\rho} = \rho \\
    & Q_t & \tilde{Q}_t = Q_t
    &&& x_0(t) & \tilde{x}_0(t) = x_0(t \tau)\\
    \multicolumn{3}{c}{}
    &&& Q_t & \tilde{Q}_t = Q_{t \tau}
\end{array}
\]
    \caption{Transformation of the transport coefficients $D(\rho)$ and $\sigma(\rho)$, of the mean density $\rho$, and of the observables (i) position of the tracer $x_0(t)$ and (ii) integrated current through the origin $Q_t$, under the different transformations identified in Sections~\ref{sec:Duality} and~\ref{sec:TrDil}. The translation of density (T) has a complex effect on the dynamics of the tracer, see Eq.~\eqref{eq:TransXtTransl}, so it is not written explicitly here.}
    \label{tab:TransfOnObs}
\end{table}

\subsection{Known results: the SEP and the KMP model}
\label{sec:KnownSEPKMP}

We summarise here the known results on the cumulant generating functions~\eqref{def:CGFs} for two models of single-file systems, the simple exclusion process and the Kipnis Marchioro Presutti model. In particular we show that the results on the KMP model can be deduced from those on the SEP.

\subsubsection{The simple exclusion process}

The simple exclusion process (SEP) is a model of particles on a lattice. Each site can host at most one particle. At random times (picked exponentially, with rate $D_0$), each particle can jump to a neighbouring site (with same probability), only if that site is empty (see Fig.~\ref{fig:Models}(a)). For this model, the transport coefficients take the form~\cite{Bodineau:2004,Derrida:2009a}
\begin{equation}
    \label{eq:TrCoefSEP}
    D(\rho) = D_0
    \:,
    \quad 
    \sigma(\rho) = 2 D_0 \rho(1-\rho)
    \:.
\end{equation}

The cumulant generating function of the integrated current in the SEP has been computed using Bethe ansatz~\cite{Derrida:2009}, and reads~\footnote{In Ref.~\cite{Derrida:2009a}, the result is given for $D_0=1$. The arbitrary value $D_0$ can be introduced by using the transformation (Rt) with $\tau = D_0$.}
\begin{equation}
    \label{eq:PsiQSEP}
    \psi_Q^{\mathrm{(SEP)}}(\lambda)
    = - \sqrt{\frac{D_0}{\pi}} \: \mathrm{Li}_{\frac{3}{2}} \left( - \omega \right)
    \:,
    \quad
    \omega = \rho (1-\rho) (\e^{\lambda} + \e^{-\lambda} - 2)
    \:,
\end{equation}
where $\mathrm{Li}_{\nu}(z) = \sum_{n \geq 1} z^n/n^\nu$ is the polylogarithm function.

The cumulant generating function of the position of the tracer has been computed more recently~\cite{Imamura:2017}, and is given as a double Legendre transform
\begin{equation}
    \label{eq:PsiTSEP}
    \psi_T^{\mathrm{(SEP)}}(s)
    = - \sqrt{D_0} \: \min_\xi \left[ 2 s \xi + \phi(\xi) \right]
    \:,
    \quad \text{where} \quad
    \phi(\xi) = \mu(\xi, \lambda^\star(\xi))
    \:,
\end{equation}
where
\begin{equation}
    \mu(\xi,\lambda) =
    \sum_{n \geq 1} \frac{(-\omega)^n}{n^{3/2}} A(\sqrt{n} \: \xi) 
    + \xi \ln \frac{1 + \rho(\e^{\lambda}-1)}{1+\rho (\e^{-\lambda}-1)}
    \:,
    \quad
    \dep{\mu}{\lambda}(\xi,\lambda^\star(\xi)) = 0
    \:,
\end{equation}
with $\omega$ defined as in~\eqref{eq:PsiQSEP} and $A(\xi) = \e^{-\xi^2}/\sqrt{\pi} + \xi \: \mathrm{erf}(\xi)$. An alternative parametric representation is given in~\cite{Grabsch:2022}. Expanding in powers of $\lambda$ gives the first cumulants (the odd order cumulants vanish by symmetry),
\begin{align}
    \moy{x_0(t)^2} 
    &=
    2\frac{1-\rho}{\rho} \sqrt{\frac{D_0 t}{\pi}}
    \:,
    \\
    \moy{x_0(t)^4}_c 
    \equiv
    \moy{x_0(t)^4} - 3 \moy{x_0(t)^2} 
    &= 2\frac{(1-\rho)}{\pi \rho^3}
    \left(
        12 (1-\rho)^2 - \pi (3 - 3 (4 - \sqrt{2})\rho + (8-3\sqrt{2}) \rho^2)
    \right)
    \sqrt{\frac{D_0 t}{\pi}}
    \:.
\end{align}
Note that, in the low density limit $\rho \to 0$, the SEP reduces to a model of Brownian particles with hard-core repulsion~\cite{Imamura:2017} (upon rescaling the distances by $\rho$ and the time by $\rho^2$), which is described by the coefficients 
\begin{equation}
    D(\rho) = D_0
    \:,
    \quad 
    \sigma(\rho) = 2 D_0 \rho
    \:.
\end{equation}
By taking the low density limit in the expressions~(\ref{eq:PsiQSEP},\ref{eq:PsiTSEP}), we obtain the cumulant generating functions for the Brownian particles. For instance, this yields~\cite{Sadhu:2015,Imamura:2017}
\begin{equation}
\label{eq:PsiTBrow}
    \psi_T^{\mathrm{(B)}}(\lambda) = 
    - \rho \sqrt{D_0} \: \min_\xi \left[ 2 \lambda \xi + \phi(\xi) \right]
    \:,
    \quad \text{where} \quad
    \phi(\xi) = \left( \sqrt{\Xi(\xi)} - \sqrt{\Xi(-\xi)} \right)^2
    \:,
    \quad 
    \Xi(\xi) = \int_\xi^{\infty} \erfc(x) \dd x
    \:.
\end{equation}

\subsubsection{A direct consequence: general quadratic mobility}

As a first illustration of the transformations identified in Sections~\ref{sec:Duality} and~\ref{sec:TrDil}, we show that we can relate any quadratic $\sigma$ (with $\sigma(0)=0$) to the SEP. This is similar to the extension discussed in Ref.~\cite{Derrida:2009a}. We consider a single-file described by the transport coefficients: 
\begin{equation}
    D(\rho) = D_0
    \:,
    \quad 
    \sigma(\rho) = 2 D_0 \rho (a - b\rho)
    \:.
\end{equation}
Applying the dilatation (Di) both for this system and its dual, we get the following expressions for the transport coefficients of the transformed systems:
\begin{equation}
\label{eq:GenSEP1}
    \begin{array}{*9{>{\displaystyle}c}}
        D_0
         &\xrightarrow[c=1/a]{\mathrm{(Di)}}
         & D_0
         &\xrightarrow{\mathrm{(Du)}}
         & \frac{D_0}{\rho^2}
         &\xrightarrow[c=1/b]{\mathrm{(Di)}}
         & \frac{D_0}{b^2 \rho^2} 
         & \xrightarrow{\mathrm{(Du)}}
         & \frac{D_0}{b^2}
         \:,
         \\[0.3cm]
         2 D_0 \rho (a - b\rho)
         &\xrightarrow[c=1/a]{\mathrm{(Di)}}
         & 2 D_0 \rho(1  - b\rho)
         &\xrightarrow{\mathrm{(Du)}}
         & 2 D_0 \left(1 - \frac{b}{\rho}\right) 
         &\xrightarrow[c=1/b]{\mathrm{(Di)}}
         & \frac{2 D_0}{b^2}\left(1 - \frac{1}{\rho} \right)
         & \xrightarrow{\mathrm{(Du)}}
         & \frac{2 D_0}{b^2}\rho (1 - \rho) \:.
    \end{array}
\end{equation}
Using finally a rescaling of time (Rt), we recover the transport coefficients of the SEP~\eqref{eq:TrCoefSEP}
\begin{equation}
\label{eq:GenSEP2}
    \begin{array}{*3{>{\displaystyle}c}}
        \frac{D_0}{b^2} 
         &\xrightarrow[\tau=b^2]{\mathrm{(Rt)}}
         & D_0 \equiv \Dt(\rho)
         \:,
         \\[0.3cm]
         \frac{2 D_0}{b^2}\rho (1 - \rho)
         &\xrightarrow[\tau=b^2]{\mathrm{(Rt)}}
         & 2 D_0 \rho(1 - \rho)
         \equiv \st(\rho) \:,
    \end{array}
\end{equation}
Under these transformations, the average density of the system becomes
\begin{equation}
\label{eq:RhoSEPgen}
        \rho
         \xrightarrow[c=1/a]{\mathrm{(Di)}}
          \frac{\rho}{a} 
         \xrightarrow{\mathrm{(Du)}}
          \frac{a}{\rho}
         \xrightarrow[c=1/b]{\mathrm{(Di)}}
          \frac{a}{b \rho} 
         \xrightarrow{\mathrm{(Du)}}
          \frac{b \rho}{a}
         \xrightarrow[\tau=b^2]{\mathrm{(Rt)}}
        \frac{b \rho}{a} \equiv \rho_{\mathrm{SEP}}
        \:,
\end{equation}
and the two observables become (on the first line, the integrated current of the transformed system is expressed in terms of the observables $Q_t$ and $x_0(t)$ of the original system; on the second line, same for the tracer's position):
\begin{equation}
\label{eq:Currentx0SEPgen}
    \begin{array}{ccccccccccc}
         Q_t
         &\xrightarrow[c=1/a]{\mathrm{(Di)}}&
         \displaystyle \frac{Q_t}{a}
         &
        \multirow{2}{*}{
        \begin{tikzpicture}
            \draw[-to] (0.,0.8) -- (0.8,0.);
            \draw[-to] (0.,0.) -- (0.8,0.8);
            \node at (0.4,0.82) {{\scriptsize(Du)}};
        \end{tikzpicture}
        }
         &
         - x_0(t)
         &\xrightarrow[c=1/b]{\mathrm{(Di)}}&
         \displaystyle - \frac{x_0(t)}{b}
         &
        \multirow{2}{*}{
        \begin{tikzpicture}
            \draw[-to] (0.,0.8) -- (0.8,0.);
            \draw[-to] (0.,0.) -- (0.8,0.8);
            \node at (0.4,0.82) {{\scriptsize(Du)}};
        \end{tikzpicture}
        }&
         \displaystyle \frac{Q_t}{a}
         &\xrightarrow[\tau=b^2]{\mathrm{(Rt)}}&
         \displaystyle \frac{Q_{tb^2}}{a}
        \equiv \tilde{Q_t} \:,
         \\
         x_0(t)
         &\xrightarrow[c=1/a]{\mathrm{(Di)}}&
          x_0(t)
         &&
         \displaystyle - \frac{Q_t}{a}
         &\xrightarrow[c=1/b]{\mathrm{(Di)}}&
         \displaystyle - \frac{Q_t}{a}
         &&
         \displaystyle \frac{x_0(t)}{b}
         &\xrightarrow[\tau=b^2]{\mathrm{(Rt)}}&
         \displaystyle \frac{x_0(tb^2)}{b}
        \equiv \tilde{x}_0(t) \:.
    \end{array}
\end{equation}
Therefore, we can relate the cumulant generating functions~\eqref{def:CGFs} by
\begin{equation}
\label{eq:PsiQGenQuad}
    \psi_Q(\lambda)
    \equiv 
    \lim_{t \to \infty} \frac{1}{\sqrt{t}} \ln \moy{\e^{\lambda Q_t}}
    = \frac{1}{b} \lim_{t \to \infty} \frac{1}{\sqrt{t/b^2}} \ln \moy{\e^{\lambda a \tilde{Q}_{t/b^2}}}
    \equiv \frac{1}{b} \: \psi_Q^{\mathrm{(SEP)}}(a \lambda)
    \:,
\end{equation}
\begin{equation}
\label{eq:PsiTGenQuad}
    \psi_T(\lambda)
    \equiv 
    \lim_{t \to \infty} \frac{1}{\sqrt{t}} \ln \moy{\e^{\lambda x_0(t)}}
    =  \frac{1}{b}
    \lim_{t \to \infty} \frac{1}{\sqrt{t/b^2}} \ln \moy{\e^{\lambda b \tilde{x}_0(t/b^2)}}
    \equiv \frac{1}{b} \: \psi_T^{\mathrm{(SEP)}}(b \lambda)
    \:,
\end{equation}
where $\psi_Q^{\mathrm{(SEP)}}$ and $\psi_T^{\mathrm{(SEP)}}$ are the cumulant generating functions of the SEP~(\ref{eq:PsiQSEP},\ref{eq:PsiTSEP}), evaluated at the density $\rho_{\mathrm{SEP}}$ given by~\eqref{eq:RhoSEPgen}. We now use these two relations to recover known results on the KMP model.

\subsubsection{The Kipnis Marchioro Presutti model}

The Kipnis Marchioro Presutti (KMP) model is a stochastic model for mass or heat transfer~\cite{Kipnis:1982}. It describes a lattice on which each site $k$ hosts a continuous variable which represents a mass (or an energy). At random times picked from an exponential distribution with rate $D_0$, a site can exchange mass with its right neighbour such that the total mass of the two neighbouring sites is randomly redistributed (with uniform distribution) between them (see Fig.~\ref{fig:Models}(d)). This model has been applied to various situations, such as force fluctuations in packs of granular beads, or distribution of wealth in a society (see Ref.~\cite{Das:2017} and references therein).

The transport coefficients of this model are given by~\cite{Zarfaty:2016}
\begin{equation}
    D(\rho) = D_0
    \:,
    \quad 
    \sigma(\rho) = \sigma_0 \rho^2
    \:.
\end{equation}
This model can be deduced from the SEP by using the transformations described in~(\ref{eq:GenSEP1},\ref{eq:GenSEP2}) by setting $\rho_\mathrm{SEP} = b \rho/a$, $b=-\sigma_0/(2D_0)$ and letting $a \to 0$. This requires to assume that the results~(\ref{eq:PsiQSEP},\ref{eq:PsiTSEP}) obtained on the SEP are valid for $\rho_\mathrm{SEP} < 0$. Under this assumption, which is also used in~\cite{Derrida:2009a}, we can deduce the cumulant generating function of the integrated current in the KMP from~\eqref{eq:PsiQGenQuad},
\begin{equation}
\label{eq:PsiQKMP}
    \psi_Q^{(\mathrm{KMP})}(\lambda) = \frac{2D_0}{\sigma_0} \sqrt{\frac{D_0}{\pi}}
    \mathrm{Li}_{\frac{3}{2}} \left( \left(\frac{\sigma_0 \rho \lambda}{2 D_0}\right)^2 \right)
    \:,
\end{equation}
given in~\cite{Derrida:2009,Derrida:2009a}.

Although the KMP is a mass transfer model and not a particle model, one can consider a tracer, at position $\tilde{x}_0(t)$ defined as in~\eqref{eq:defx0}. Its position can be interpreted as a fictitious "wall" that separates the systems into two regions in which the mass is conserved. Applying the same method as for the current, the corresponding cumulant generating function $\psi_T^{\mathrm{(KMP)}}$ can be deduced from the one of the SEP using~\eqref{eq:PsiTGenQuad}, and reads,
\begin{equation}
    \label{eq:PsiTKMP}
    \psi_T^{\mathrm{(KMP)}}(\lambda)
    =
    -\frac{2 D_0}{\sigma_0}
    \lim_{a \to 0}
    \left.
    \psi_T^{\mathrm{(SEP)}} \left( - \frac{\sigma_0}{2 D_0} \lambda \right)
    \right|_{\rho_{\mathrm{SEP}=-\sigma_0 \rho/(2 D_0 a)}}
    \:.
\end{equation}
The limit is not easily taken on the expressions~\eqref{eq:PsiTSEP}, but this can be done straightforwardly on the cumulants. For instance, this gives
\begin{equation}
\label{eq:CumulTKMP}
    \moy{x_0(t)^2} = \sigma_0 \sqrt{\frac{t}{D_0 \pi}}
    \:,
    \qquad
    \moy{x_0(t)^4}_c
    = \sigma_0^3 \frac{12+ \pi(3 \sqrt{2}-8)}{4 D_0^2 \pi}
    \sqrt{\frac{t}{D_0 \pi}}
    \:.
\end{equation}
An alternative parametric representation of $\psi_T^{\mathrm{(KMP)}}(\lambda)$ is given in~\cite{Grabsch:2022}.

\subsection{The random average process}

The random average process (RAP)~\cite{Ferrari:1998,Krug:2000,Rajesh:2001} is a system of particles placed on an infinite line.
At random times, picked from an exponential distribution with rate $\frac{1}{2}$, a particle can move, either to the left or to the right, to a random fraction of the distance to the next particle (see Fig.~\ref{fig:Models}(c)). In the hydrodynamic limit, only the first two moments $\mu_1$ and $\mu_2$ of the distribution of this random fraction are relevant. The transport coefficients depend on these two parameters only~\cite{Krug:2000,Kundu:2016}:
\begin{equation}
    D(\rho) = \frac{\mu_1}{2\rho^2}
    \:,
    \quad 
    \sigma(\rho) = \frac{1}{\rho} \frac{ \mu_1 \mu_2}{\mu_1 - \mu_2}
    \:.
\end{equation}
Applying the duality relation~\eqref{eq:RelCoefsTransp}, we obtain that the dual system is described by the coefficients
\begin{equation}
\label{eq:CoefsDualRAP}
    \Dt(\rt) = D_0 = \frac{\mu_1}{2}
    \:,
    \quad 
    \st(\rt) = \sigma_0 \rt^2
    \:,
    \quad
    \sigma_0 = \frac{ \mu_1 \mu_2}{\mu_1 - \mu_2}
    \:,
\end{equation}
which correspond to the KMP model, with a specific choice of the factors $D_0$ and $\sigma_0$ in the coefficients (see Table~\ref{tab:TransportCoefs}). Note that, under this duality transformation, the mean density $\rho$ of the RAP becomes
\begin{equation}
    \label{eq:TrAvgDens}
    \rho_{\mathrm{KMP}} = \frac{1}{\rho} 
\end{equation}
for the dual system.

\subsubsection{Integrated current through the origin}

The integrated current through the origin $Q_t$ is related to the position of the tracer in the dual system $\tilde{x}_0(t)$ via~\eqref{eq:x0DualFromQ}. This implies that the cumulant generating function of $Q_t$ reads
\begin{equation}
    \psi_Q^{\mathrm{(RAP)}}(\lambda)
    \equiv 
    \lim_{t \to \infty} \frac{1}{\sqrt{t}} \ln \moy{\e^{\lambda Q_t}}
    = \lim_{t \to \infty} \frac{1}{\sqrt{t}} \ln \moy{\e^{-\lambda \tilde{x}_0(t)}}
    \equiv \psi_T^{\mathrm{(KMP)}}(-\lambda)
    \:,
\end{equation}
where $\psi_T^{\mathrm{(KMP)}}$ is the cumulant generating function of a tracer in the KMP model~\eqref{eq:PsiTKMP}, at density $\rho_{\mathrm{KMP}}$ given by~\eqref{eq:TrAvgDens}. In particular, the first cumulants read,
\begin{equation}
    \moy{Q_t^2} = \sigma_0 \sqrt{ \frac{t}{D_0 \pi} }
    \:,
    \quad
    \moy{Q_t^4}_c \equiv 
    \moy{Q_t^4} - 3 \moy{Q_t^2}
    =
    \sigma_0^3 \frac{12+ \pi(3 \sqrt{2}-8)}{4 D_0^2 \pi}
    \sqrt{\frac{t}{D_0 \pi}}
    \:.
\end{equation}
Remarkably, these cumulants (and also the cumulant generating function) do not depend on the mean density $\rho$ of the RAP. This is due to the fact that the dynamics of the RAP is invariant under a rescaling of the positions of all the particles (which does not change the integrated current $Q_t$).

\subsubsection{Position of a tracer}

The position $x_0(t)$ of the tracer in the RAP can be expressed in terms of the flux of the dual system (the KMP model) using~\eqref{eq:x0fromQdual}. This relation was used in Refs.~\cite{Cividini:2016a,Cividini:2016,Kundu:2016} to study the fluctuations of the position of a tracer, and correlations between different particles in the RAP. Here, we use it to express the generating function of the cumulants of $x_0(t)$ as
\begin{equation}
    \psi_T^{\mathrm{(RAP)}}(\lambda) \equiv 
    \lim_{t \to \infty} \frac{1}{\sqrt{t}} \ln \moy{\e^{\lambda x_0(t)}}
    = \lim_{t \to \infty} \frac{1}{\sqrt{t}} \ln \moy{\e^{-\lambda \tilde{Q}_t}}
    \equiv \psi_Q^{\mathrm{(KMP)}}(-\lambda)
    \:,
\end{equation}
hence, from~\eqref{eq:PsiQKMP},
\begin{equation}
    \psi_T^{(\mathrm{RAP})}(\lambda) = 
    \frac{(\mu_1-\mu_2)}{\mu_2}
    \sqrt{\frac{\mu_1}{2\pi}}
    \mathrm{Li}_{\frac{3}{2}} \left( \left(\frac{\mu_2 \lambda}{\rho(\mu_1-\mu_2)}\right)^2 \right)
    \:.
\end{equation}
Expanding this expression in powers of $\lambda$, we obtain the cumulants of the position of the tracer. For instance,
\begin{equation}
    \moy{x_0(t)^2} = \frac{\mu_2}{(\mu_1-\mu_2)\rho^2} \sqrt{\frac{2\mu_1 t}{\pi}}
    \:,
    \qquad
    \moy{x_0(t)^4}_c \equiv
    \moy{x_0(t)^4} - 3 \moy{x_0(t)^2} 
    = \frac{6 \mu_2^3}{(\mu_1-\mu_2)^3 \rho^4} \sqrt{\frac{\mu_1 t}{\pi}}
    \:.
\end{equation}
The variance coincides with the known result~\cite{Rajesh:2001}.

\subsection{The zero range process}

Zero range processes (ZRP)~\cite{Evans:2005} correspond to lattice models, in which each site $k$ can be occupied by any number $n_k$ of particles. At random times picked from an exponential distribution, each occupied site can lose a particle, which hops either to the left or to the right with equal probability (see Fig.~\ref{fig:Models}(b)). The hopping rates from a site $k$ can depend on the occupation number $n_k$ of the site, but not on the occupations of the other sites (unlike exclusion processes), hence the name "zero range". At the macroscopic scale, they correspond to a wide class of transport coefficients (depending on the hopping rates), such that they are related by $D(\rho) = \sigma'(\rho)/2$~\cite{Bodineau:2004}.

Here, we consider one specific ZRP, which corresponds to the simplest case in which the hoppings are constant, so they depend neither on the site nor on the number of particles on it (note that this is different from a system of independent particles, as in this case the hopping rate on site $k$ would be proportional to $n_k$). At the macroscopic level, this system is described by the transport coefficients~\cite{Landim:1998,Bodineau:2004}
\begin{equation}
    D(\rho) = \frac{D_0}{(1+\rho)^2}
    \:,
    \qquad
    \sigma(\rho) = \frac{2 D_0 \rho}{1+\rho} 
    \:.
\end{equation}
This model is remarkable because it is an example of system whose transport coefficients are left unchanged by the duality transformation (Du). This property will be useful below.
This model is well-known to be related to the SEP~\cite{Evans:2000,Evans:2005}, and this relation was used explicitly in~\cite{Landim:1998}. The mapping is the following: to each site $k$ is associated a particle at position $x_k$. The occupation of site $k$ in the ZRP is given by the number of empty sites in the SEP between particles $k+1$ and $k$ (see Fig.~\ref{fig:MappingSEPtoZRP}). Inversely, the distance between particles $k+1$ and $k$ is given by $n_k+1$. Therefore, the SEP can be obtained from the ZRP by combining the translation of density (T) and the duality relation (Du):
\begin{equation}
    \begin{array}{*5{>{\displaystyle}c}}
        D(\rho) = \frac{D_0}{(1+ \rho)^2}
         &\xrightarrow[c=1]{\mathrm{(T)}}
         &\frac{D_0}{\rho^2}
         &\xrightarrow{\mathrm{(Du)}}
         & D_0= \Dt(\rho) \:,
         \\[0.3cm]
         \sigma(\rho) = \frac{2D_0 \rho}{1+\rho} 
         &\xrightarrow[c=1]{\mathrm{(T)}}
         & \frac{2 D_0}{\rho}(\rho-1)
         &\xrightarrow{\mathrm{(Du)}}
         & 2 D_0 \rho (1-\rho) = \st(\rho) \:.
    \end{array}
\end{equation}
Under theses transformations the observables become (again, the first line represents the current in the transformed system in terms of the observables of the original system; the second line is the expression of the position of the tracer):
\begin{equation}
\label{eq:RelFluxZRP}
    \begin{array}{ccccccccccc}
         Q_t
         &\xrightarrow{\mathrm{(T)}}&
         \displaystyle Q_t
         &
        \multirow{2}{*}{
        \begin{tikzpicture}
            \draw[-to] (0.,0.8) -- (0.8,0.);
            \draw[-to] (0.,0.) -- (0.8,0.8);
            \node at (0.4,0.82) {{\scriptsize(Du)}};
        \end{tikzpicture}
        }
         & - \mathcal{X}_t[x_0,\rho] \equiv \tilde{Q}_t
         \:,
         \\[0.4cm]
         x_0(t)
         &\xrightarrow{\mathrm{(T)}}&
         \mathcal{X}_t[x_0,\rho]
         &&
         -Q_t \equiv \tilde{x}_0(t)
         \:,
    \end{array}
\end{equation}
where we denoted $\mathcal{X}_t[x_0,\rho]$ the result of the complex transformation of the position of the tracer~\eqref{eq:TransXtTransl} under (T).
The mean density of the ZRP is transformed as
\begin{equation}
\label{eq:DensZRPSEP}
    \begin{array}{*5{>{\displaystyle}c}}
        \rho
         &\xrightarrow[c=1]{\mathrm{(T)}}
         & \rho + 1
         &\xrightarrow{\mathrm{(Du)}}
         & \frac{1}{\rho+1} = \rho_{\mathrm{SEP}}
         \:.
    \end{array}
\end{equation}
Using~\eqref{eq:RelFluxZRP}, we can directly obtain the cumulant generating function for the current
\begin{equation}
    \psi_Q^{\mathrm{(ZRP)}}(\lambda) \equiv 
    \lim_{t \to \infty} \frac{1}{\sqrt{t}} \ln \moy{\e^{\lambda Q_t}}
    = \lim_{t \to \infty} \frac{1}{\sqrt{t}} \ln \moy{\e^{-\lambda \tilde{x}_0(t)}}
    \equiv \psi_T^{\mathrm{(SEP)}}(-\lambda)
    \:,
\end{equation}
where $\psi_T^{\mathrm{(SEP)}}$ is given by~\eqref{eq:PsiTSEP}, evaluated at the density $\rho_{\mathrm{SEP}}$~\eqref{eq:DensZRPSEP}. For instance,
\begin{equation}
    \label{eq:ExampleQZRP}
    \moy{Q_t^2} = 2 \rho \sqrt{\frac{D_0 t}{\pi}}
    \:,
    \quad
    \moy{Q_t^4}_c
    = 2 \rho \frac{12 \rho^2 + \pi (1+3(2-\sqrt{2})\rho -3 \rho^2)}{\pi}
    \sqrt{\frac{D_0 t}{\pi}}
    \:.
\end{equation}

In order to obtain the cumulant generating function for the tracer in the ZRP, one should in principle use the complex transformation $\mathcal{X}_t[x_0,\rho]$, which requires the knowledge of the statistical properties of the density, as well as its correlations with the position of the tracer. These correlations are however more complex objects than the cumulant generating functions~\cite{Poncet:2021}, and are not known for the ZRP. However, here, we can circumvent this problem since the ZRP we consider is self dual:
\begin{equation}
    \begin{array}{*5{>{\displaystyle}c}}
        D(\rho) = \frac{D_0}{(1+ \rho)^2}
         &\xrightarrow{\mathrm{(Du)}}
         & \frac{D_0}{(1+ \rho)^2} = D(\rho) \:,
         \\[0.3cm]
         \sigma(\rho) = \frac{2D_0 \rho}{1+\rho} 
         &\xrightarrow{\mathrm{(Du)}}
         & \frac{2D_0 \rho}{1+\rho}  = \sigma(\rho) \:.
    \end{array}
\end{equation}
The corresponding action on the observables is an exchange of integrated current and flux : 
\begin{equation}
    \begin{array}{cccc}
         Q_t
         
         &
        \multirow{2}{*}{
        \begin{tikzpicture}
            \draw[-to] (0.,0.8) -- (0.8,0.);
            \draw[-to] (0.,0.) -- (0.8,0.8);
            \node at (0.4,0.82) {{\scriptsize(Du)}};
        \end{tikzpicture}
        }
         & - x_0(t) \equiv \tilde{Q}_t
         \:,
         \\[0.4cm]
         x_0(t)
         &
         &
         -Q_t \equiv \tilde{x}_0(t)
         \:,
    \end{array}
\end{equation}
and the initial mean density is mapped in the following way : 
\begin{equation}
    \begin{array}{*5{>{\displaystyle}c}}
        \rho
         &\xrightarrow{\mathrm{(Du)}}
         & \frac{1}{\rho} = \tilde{\rho}
         \:.
    \end{array}
\end{equation}
This mapping thus implies a direct relation of the cumulant generating functions for the tracer and the current in the ZRP:
\begin{equation}
    \psi_T^{\mathrm{(ZRP)}}(\lambda, \rho)
    = \psi_Q^{\mathrm{(ZRP)}}(-\lambda, \rho^{-1})
    \:,
\end{equation}
from which we can deduce for instance, using~\eqref{eq:ExampleQZRP},
\begin{equation}
    \moy{x_0(t)^2} = \frac{2}{\rho} \sqrt{\frac{D_0 t}{\pi}}
    \:,
    \quad
    \moy{x_0(t)^4}_c
    = 2 \frac{12  + \pi (\rho^2+3(2-\sqrt{2})\rho -3 )}{\rho^3 \pi}
    \sqrt{\frac{D_0 t}{\pi}}
    \:.
\end{equation}

\subsection{The gas of hard rods}

Consider a system of rods of length $\ell$, which perform a Brownian motion, with the condition that two rods cannot overlap (see Fig.~\ref{fig:Models}(f)). The main interest of this model compared to the point-like Brownian particles is that the system of hard rods has a maximal density $1/\ell$. It is the one dimensional version of the hard sphere gas. Its transport coefficients are given by~\cite{Lin:2005}
\begin{equation}
    D(\rho) =\frac{D_0}{(1-\ell \rho)^2}
    \:,
    \qquad
    \sigma(\rho) = 2 D_0 \rho
    \:.
\end{equation}

We can map the gas of hard rods onto the model of point-like Brownian particles by decreasing the size of the rods by their length $\ell$. This is naturally done in the dual model (which describes the gaps between the rods) by removing a constant length $\ell$ to all the gaps. We can therefore construct this model by going to the dual of the hard rods model, translating the density by a constant $-\ell$ and going back to the particles using the duality relation again:
\begin{equation}
    \begin{array}{*7{>{\displaystyle}c}}
        D(\rho) = \frac{D_0}{(1-\ell \rho)^2}
         &\xrightarrow{\mathrm{(Du)}}
         &\frac{D_0}{(\rho-\ell)^2}
         &\xrightarrow[c=-\ell]{\mathrm{(T)}}
         &\frac{D_0}{\rho^2}
         &\xrightarrow{\mathrm{(Du)}}
         &D_0 = \Dt(\rho) \:,
         \\[0.3cm]
         \sigma(\rho) = 2D_0 \rho 
         &\xrightarrow{\mathrm{(Du)}}
         & 2 D_0
         &\xrightarrow[c=-\ell]{\mathrm{(T)}}
         & 2 D_0
         &\xrightarrow{\mathrm{(Du)}}
         & 2 D_0 \rho = \st(\rho) \:.
    \end{array}
\end{equation}
$D(\rho)$ and $\sigma(\rho)$ are the transport coefficients for the gas of hard rods~\cite{Lin:2005}, and this series of transformation indeed yields the coefficients $\Dt$ and $\st$ of the gas of Brownian particles (see Table~\ref{tab:TransportCoefs}). Under these transformations (see Table~\ref{tab:TransfOnObs}), the position of the tracer is unchanged (in each column we give the observables of the current system in terms of the observables of the original system; the first line corresponds to the integrated current, the second line is the position of the tracer):
\begin{equation}
    \begin{array}{ccccccc}
         Q_t
         &\multirow{2}{*}{
        \begin{tikzpicture}
            \draw[-to] (0.,0.8) -- (0.8,0.);
            \draw[-to] (0.,0.) -- (0.8,0.8);
            \node at (0.4,0.82) {{\scriptsize(Du)}};
        \end{tikzpicture}
        }&
         \displaystyle -x_0(t)
         &\xrightarrow[c=-\ell]{\mathrm{(T)}}
         & -x_0(t)
         &\multirow{2}{*}{
        \begin{tikzpicture}
            \draw[-to] (0.,0.8) -- (0.8,0.);
            \draw[-to] (0.,0.) -- (0.8,0.8);
            \node at (0.4,0.82) {{\scriptsize(Du)}};
        \end{tikzpicture}
        }&
        -? \equiv \tilde{Q}_t
         \:,
         \\[0.4cm]
         x_0(t)
         &&
         -Q_t
         &\xrightarrow[c=-\ell]{\mathrm{(T)}}&
         ?
         &&
         x_0(t) \equiv \tilde{x}_0(t)
         \:,
    \end{array}
\end{equation}
where we again do not write explicitly the complex transformation~\eqref{eq:TransXtTransl} of the position of the tracer under the transformation (T). The mean density of the system becomes
\begin{equation}
\label{eq:DensHRB}
    \begin{array}{*7{>{\displaystyle}c}}
         \rho
         &\xrightarrow{\mathrm{(Du)}}
         & \frac{1}{\rho}
         &\xrightarrow[c=-\ell]{\mathrm{(T)}}
         & \frac{1}{\rho} - \ell
         &\xrightarrow{\mathrm{(Du)}}
         & \frac{\rho }{1 - \rho  \ell} \equiv \rho_\mathrm{B}
         \:.
    \end{array}
\end{equation}
Since the transformation of the flux is rather complex, and requires the tracer/density correlations in the dual system (which are not known), we will only consider the cumulant generating function of the tracer, which reads
\begin{equation}
    \psi_T^{\mathrm{(HR)}}(\lambda) \equiv 
    \lim_{t \to \infty} \frac{1}{\sqrt{t}} \ln \moy{\e^{\lambda x_0(t)}}
    = \lim_{t \to \infty} \frac{1}{\sqrt{t}} \ln \moy{\e^{\lambda \tilde{x}_0(t)}}
    \equiv \psi_T^{\mathrm{(B)}}(\lambda)
    \:,
\end{equation}
in terms of the cumulant generating function $\psi_T^{\mathrm{(B)}}$ of the tracer in the Brownian model~\eqref{eq:PsiTBrow}, evaluated at the density $\rho_B$~\eqref{eq:DensHRB}.
In particular, expanding in powers of $\lambda$, we get the first cumulants:
\begin{equation}
    \moy{x_0(t)^2} = 
    \frac{2 (1-\ell \rho)}{\rho}
    \sqrt{\frac{D_0 t}{\pi}}
    \:,
    \qquad
    \moy{x_0(t)^4}_c 
    = \frac{6(4-\pi)}{\pi} \left(\frac{1- \ell \rho}{\rho}\right)^3
    \sqrt{\frac{D_0 t}{\pi}}
    \:.
\end{equation}

\subsection{The double exclusion process}

We consider an exclusion process, with additional repulsion between neighbouring sites: a particle located at site $i$ can, at exponential times, jump to site $i+1$ only if sites $i+1$ and $i+2$ are empty, or with the same probability to site $i-1$ only if sites $i-1$ and $i-2$ are empty. This model is equivalent to having particles which occupy a volume of two lattice sites, with the constraint that they cannot overlap (see Fig.~\ref{fig:Models}(e)). Exclusion processes with extended particles have for instance been considered for applications to biological systems~\cite{Shaw:2003}.
The transport coefficients are~\cite{Hager:2001,Baek:2017}
\begin{equation}
     D(\rho) = \frac{D_0}{(1-\rho)^2}
     \:,
     \qquad
     \sigma(\rho) = 2 D_0 \frac{\rho(1-2\rho)}{1-\rho} 
     \:.
\end{equation}

This model can be directly mapped onto the SEP by reducing the size of the particles to one lattice site, similarly as we did for the gas of hard rods:
\begin{equation}
    \begin{array}{*7{>{\displaystyle}c}}
        D(\rho) = \frac{D_0}{(1-\rho)^2}
         &\xrightarrow{\mathrm{(Du)}}
         &\frac{D_0}{(\rho-1)^2}
         &\xrightarrow[c=-1]{\mathrm{(T)}}
         &\frac{D_0}{\rho^2}
         &\xrightarrow{\mathrm{(Du)}}
         &D_0 = \Dt(\rho) \:,
         \\[0.3cm]
         \sigma(\rho) = 2 D_0 \frac{\rho(1-2\rho)}{1-\rho} 
         &\xrightarrow{\mathrm{(Du)}}
         & 2 D_0 \left(1 - \frac{1}{\rho-1} \right) 
         &\xrightarrow[c=-1]{\mathrm{(T)}}
         & 2 D_0 \left(1 - \frac{1}{\rho} \right)
         &\xrightarrow{\mathrm{(Du)}}
         & 2D_0 \rho  (1-\rho) = \st(\rho) \:,
    \end{array}
\end{equation}
which are indeed the transport coefficients of the SEP (see Table~\ref{tab:TransportCoefs}). Note that we can straightforwardly extend this procedure to a system in which particles have a volume on $n$ sites, by applying a translation of density (T) of $n-1$ for the dual. Under this series of transformations, the position of a tracer becomes (we again do not write the action of (T) on the position of the tracer of the dual system because of its complexity)
\begin{equation}
    \begin{array}{ccccccc}
         Q_t
         &\multirow{2}{*}{
        \begin{tikzpicture}
            \draw[-to] (0.,0.8) -- (0.8,0.);
            \draw[-to] (0.,0.) -- (0.8,0.8);
            \node at (0.4,0.82) {{\scriptsize(Du)}};
        \end{tikzpicture}
        }&
         \displaystyle -x_0(t)
         &\xrightarrow[c=-1]{\mathrm{(T)}}
         & -x_0(t)
         &\multirow{2}{*}{
        \begin{tikzpicture}
            \draw[-to] (0.,0.8) -- (0.8,0.);
            \draw[-to] (0.,0.) -- (0.8,0.8);
            \node at (0.4,0.82) {{\scriptsize(Du)}};
        \end{tikzpicture}
        }&
        -? \equiv \tilde{Q}_t
         \:,
         \\[0.4cm]
         x_0(t)
         &&
         -Q_t
         &\xrightarrow[c=-1]{\mathrm{(T)}}&
         ?
         &&
         x_0(t) \equiv \tilde{x}_0(t)
         \:,
    \end{array}
\end{equation}
and the density
\begin{equation}
\label{eq:DensDEP}
    \begin{array}{*7{>{\displaystyle}c}}
         \rho  
         &\xrightarrow{\mathrm{(Du)}}
         & \frac{1}{\rho}
         &\xrightarrow[c=-1]{\mathrm{(T)}}
         & \frac{1}{\rho} - 1
         &\xrightarrow{\mathrm{(Du)}}
         & \frac{\rho}{1 - \rho} \equiv \rho_\mathrm{SEP}
         \:.
    \end{array}
\end{equation}
The cumulant generating function of the tracer in the double exclusion process is thus
\begin{equation}
    \psi_T^{\mathrm{(DEP)}}(\lambda) \equiv 
    \lim_{t \to \infty} \frac{1}{\sqrt{t}} \ln \moy{\e^{\lambda x_0(t)}}
    = \lim_{t \to \infty} \frac{1}{\sqrt{t}} \ln \moy{\e^{\lambda \tilde{x}_0(t)}}
    \equiv \psi_T^{\mathrm{(SEP)}}(\lambda)
    \:,
\end{equation}
where $\psi_T^{\mathrm{(SEP)}}$ is given by~\eqref{eq:PsiTSEP}, evaluated at density $\rho_\mathrm{SEP}$ given by~\eqref{eq:DensDEP}. In particular, expanding in powers of $\lambda$, we get the first cumulants:
\begin{align}
    \moy{x_0(t)^2} 
    &= 
    \frac{2(1-2\rho)}{\rho} \sqrt{\frac{D_0 t}{\pi}}
    \:,
    \\
    \moy{x_0(t)^4}_c 
    &= \frac{2(1-2\rho)}{\pi \rho^3}
    \left( 12 (1-2 \rho )^2-\pi  ((23-6 \sqrt{2}) \rho^2-3
   (6-\sqrt{2}) \rho +3) \right)
    \sqrt{\frac{D_0 t}{\pi}}
    \:.
\end{align}

\section{Conclusion}

We have shown that, in the framework of fluctuating hydrodynamics, the duality relation that is well known between the SEP and the ZRP, and between the RAP and the KMP model, is not restricted to these models. Any single-file system can be mapped onto a dual one, which still obeys the equations of fluctuating hydrodynamics, with new transport coefficients.
We have also introduced three other transformations, which, combined with the duality relation, yield all the possible relations between two single-file models.

We have characterized the action of these transformations on the two main observables studied in single-file diffusion: the integrated current through the origin and the position of a tagged particle. As an application of these mappings, we have shown that the cumulant generating functions of these observables can be obtained for various models of single-file systems from the one computed for the SEP.

Dualities have been essential in the resolution of microscopic models such as the SEP, the KMP model or the RAP~\cite{Kipnis:1982,Landim:1998,Bertini:2005,Carinci:2013,Cividini:2016a}. The mappings presented here, valid for any single-file system, could play an important role in the application of the MFT to new models of single-file diffusion.

\appendix

\section{General mapping between two single-file systems}
\label{sec:AppOnlyMappings}

Here we investigate the possibility of existence of other mappings than the physical ones presented in the main text. We look for transformations of the following general form, linear with respect to the flux (the case of a non linear transformation of the flux generates difficulties concerning the random white noise):  
\begin{align}
    \rt(k,t) & = F(\rho(x_t(k),t/\tau))
    \:,
\label{eq:rtgen}\\
    \jt(k,t) & = j(x_t(k),t/\tau) G(\rho(x_t(k),t/\tau))
    \:,
\label{eq:jtgen}\\
    \partial_k x_t(k) & = 
    \dfrac{F'(\rho(x_t(k'),t/\tau))}{\tau G(\rho(x_t(k'),t/\tau))}
    \:,
\label{eq:xkgen}\\
    \partial_t x_t(k) & = 
    -j(x_t(k),t/\tau)\dfrac{G'(\rho(x_t(k),t/\tau))}{\tau G(\rho(x_t(k),t/\tau))}
    \:.
\label{eq:xtgen}
\end{align}
The definitions \eqref{eq:xkgen}, \eqref{eq:xtgen} are necessary so that the new density and flux verify the conservation equation
\begin{equation}
    \partial_t \rt(k,t) + \partial_k \jt(k,t) = 0    
    \:.
\end{equation}
Then, from \eqref{eq:StochCurrent}, we get the following stochastic Fourier law
\begin{equation}
    \jt(k,t) + \Dt (\rt)\partial_k \rt(k,t) = \sqrt{\st(\rt)} \eta_{t,k}
    \:,
\end{equation}
where
\begin{align}
\Dt(x) &  = \tau\left(\dfrac{G^2}{F'^2} D \right) \circ F^{-1}(x)
    \:,\\
\st(x) &  = \tau^2\left(\dfrac{\lvert G \rvert ^3}{F'}\sigma \right) \circ F^{-1}(x)
    \:.
\end{align}
This mapping is not coherent for general $F$ and $G$ because of the compatibility condition $\partial_t\partial_k x_t(k) = \partial_k \partial_t x_t(k)$ which implies
\begin{align}
\left( - \dfrac{j G'\circ\rho}{G\circ\rho}\partial_x\rho +\partial_t \rho\right) \left(\dfrac{F''\circ\rho}{G\circ\rho}
- \dfrac{F'\circ\rho G'\circ\rho}{(G\circ\rho)^2}\right) 
=
 & -\dfrac{F'\circ\rho}{(G\circ\rho)^2}(-\partial_t \rho G'\circ\rho + j \partial_x \rho G''\circ\rho) \nonumber\\
& + \dfrac{(F'\circ\rho) (G'\circ\rho)^2}{(G\circ\rho)^3}j \partial_x \rho
\:.
\end{align} 
The factors in front of $j \partial_x \rho$ and $\partial_t \rho$ must vanish, so we get after simplifications,
\begin{align}
    G'F'' & = F'G''
    \:,\\
    2\dfrac{F'G'}{G} & = F'' 
    \:.
\end{align}
The only solutions to these equations are parametrized by $a, b, c, d \in \mathbb{R}$: 
\begin{align}
    G(x) = & \dfrac{1}{a + b x} \:,
    \label{eq:mapG} \\
    F(x) = & \left\lvert \begin{matrix}
    c + \dfrac{d}{a + b x} && ~~~~~~~~\text{if } b \neq 0  \:,\\
    c + d x         && ~~~~~~~~\text{if } b = 0 \:. \\
\end{matrix}\right. \label{eq:mapF}
\end{align}
The particular case $a = 0, b = -1, c = 0, d = -1, \tau = 1$ corresponds to the particles/gaps duality relation (Du). The transformation (T) of the main text is recovered by taking $a = 1, b = 0, d = 1, \tau = 1$ and (Di) by taking $a = 1/d, b = 0, c = 0, \tau = 1$. The time rescaling (Rt) is the case $a = \tau, b = 0, c = 0, d = 1$. Conversely, by composing the four transformations, we can reconstruct any possible transformation of the form~(\ref{eq:rtgen},\ref{eq:jtgen},\ref{eq:xkgen},\ref{eq:xtgen}). For the case $b=0$ in \eqref{eq:mapF}, one can perform (in this order) (Du), (Di), (Du), (Di), (T), (Rt) and for the other case, (Di), (Du), (Di), (T), (Du), (T), (Rt) generates all the possibilities.

\section{Mapping of the MFT equations}
\label{sec:AppMFT}

In the main text, we have used the formalism of fluctuating hydrodynamics, in which the density field $\rho$ and the current $j$ are stochastic. In practice, it is convenient to use the Macroscopic Fluctuation Theory (MFT), which is a deterministic rewriting of the fluctuating hydrodynamics~\cite{Bertini:2001,Bertini:2002,Bertini:2009,Bertini:2015}. We briefly recall the main steps of this process, as presented in~\cite{Derrida:2009a}. The idea is to start from the distribution of the white noise $\eta(x,t)$,
\begin{equation}
    P[\eta] \propto \exp \left[ 
    - \int_{0}^T \dd t \int_{-\infty}^\infty \dd x \: \frac{\eta(x,t)^2}{2}
    \right]
    \:,
\end{equation}
and insert the relation~\eqref{eq:StochCurrent} in order to get the probability of observing $\rho$ and $j$ at all positions and times between $0$ and $T$:
\begin{equation}
\label{eq:JointDistrRhoJ}
    P[\{ \rho(x,t), j(x,t) \}] \propto \exp \left[ 
    - \int_{0}^T \dd t \int_{-\infty}^\infty \dd x \:
    \frac{[j(x,t) + D(\rho(x,t)) \partial_x \rho(x,t)]^2}{2 \sigma(\rho(x,t))^2}
    \right]
    \: \delta \left( \partial_t \rho + \partial_x j \right)
    \:,
\end{equation}
where the $\delta$-function imposes the continuity relation~\eqref{eq:ContRel} at all points $(x,t)$. This constraint can be handled by using the Martin-Siggia-Rose formalism, which consists in replacing the $\delta$ by the integration over a conjugate field $H(x,t)$,
\begin{equation}
    \delta \left( \partial_t \rho + \partial_x j \right)
    = \int \D H \: \exp \left[ -  \int_{0}^T \dd t \int_{-\infty}^\infty \dd x \: H(\partial_t \rho + \partial_x j)   \right]
    \:.
\end{equation}
Inserting this relation in~\eqref{eq:JointDistrRhoJ} and integrating over $j$, we obtain the probability of observing a realisation $\rho(x,t)$,
\begin{multline}
\label{eq:DistrRho0}
    P[\{ \rho(x,t) \}] \propto \int \D H \D j \exp \left\lbrace
    - \int_{0}^T \dd t \int_{-\infty}^\infty \dd x \: \left[
    H \partial_t \rho + D(\rho) \partial_x \rho \partial_x H
    - \frac{\sigma(\rho)}{2} (\partial_x H)^2
    \right.
    \right.
    \\
    \left.
    \left.
    + \frac{[j + D(\rho) \partial_x \rho - \sigma(\rho) \partial_x H]^2}{2 \sigma(\rho)^2}
    \right]
    \right\rbrace
    \:,
\end{multline}
after integration by parts. The integral over $j$ is Gaussian and can thus be performed straightforwardly,
\begin{equation}
    P[\{ \rho(x,t) \}] \propto \int \D H \exp \left\lbrace
    - \int_{0}^T \dd t \int_{-\infty}^\infty \dd x \: \left[
    H \partial_t \rho + D(\rho) \partial_x \rho \partial_x H
    - \frac{\sigma(\rho)}{2} (\partial_x H)^2
    \right]
    \right\rbrace
    \:.
\end{equation}
Rescaling the time $t \to t/T$ and space $x \to x/\sqrt{T}$, one can see that the argument of the exponential is proportional to $\sqrt{T}$. Therefore, for large $T$, the integral can be estimated by the saddle point method. Denoting $(q,p)$ the values of $(\rho,H)$ which dominate the integral, one finds that they obey the MFT equations:
\begin{align}
\label{eq:MFTq}
    \partial_t q 
    &=
    \partial_x (D(q)\partial_x q)
    - \partial_x (\sigma(q) \partial_x p)
    \:,
    \\
\label{eq:MFTp}
    \partial_t p
    &= -D(q) \partial_x^2 p
    - \frac{\sigma'(q)}{2} (\partial_x p)^2
    \:.
\end{align}
Note that the Gaussian integral over $j$ in~\eqref{eq:DistrRho0} can also be computed by a saddle point method, and this gives the optimal current
\begin{equation}
    j_\star = - D(q) \partial_x q + \sigma(q) \partial_x p
    \:.
\end{equation}
With this expression, the first MFT equation~\eqref{eq:MFTq} can be interpreted as the conservation relation~\eqref{eq:ContRel} for the optimal density $q$ and optimal current $j_\star$.

We now study the consequences of the duality relation discussed in Section~\ref{sec:Duality} on the MFT equations~(\ref{eq:MFTq},\ref{eq:MFTp}). Relations~(\ref{eq:DualFields},\ref{eq:DualPos}) hold for any realisation of the fluctuating fields $\rho$ and $j$. They thus remain valid for their typical realisation $q$ and $j_\star$. This allows to define a dual optimal field
\begin{equation}
\label{eq:MFTqDual}
    \qt(k(x,t),t) = \frac{1}{q(x,t)}
    \:,
    \quad
    k(x,t) = \int_0^x q(x',t) \dd x' + k(0,t)
    \:,
    \quad
    \partial_t k(x,t) = -j_\star(x,t)
    \:,
\end{equation}
and a dual optimal current,
\begin{equation}
    \jt_\star(k,t) = - \frac{j_\star(x,t)}{q(x,t)}
    = - \qt(k,t) j_\star(x,t)
    = - \Dt(\qt) \partial_k \qt - \st(\qt) \partial_x p
    \:,
\end{equation}
with $\Dt$ and $\st$ given by~\eqref{eq:RelCoefsTransp}.
We define $\pt(k,t)$ such that
\begin{equation}
\label{eq:RelPdual}
    \partial_k \pt(k,t) = -\partial_x p (x_k(t),t)
    \:,
\end{equation}
so that the dual current reads 
\begin{equation}
    \jt_\star = - \Dt(\qt) \partial_k \qt + \sigma(\qt) \partial_k \pt
    \:.
\end{equation}
Using relations~(\ref{eq:dtrho},\ref{eq:dxrho},\ref{eq:dxj}), applied to the optimal fields, we then deduce
\begin{equation}
    \partial_t \qt
    =
    \partial_k (\Dt(\qt)\partial_k \qt)
    - \partial_k (\sigma(\qt) \partial_k \pt)
    \:,
\end{equation}
which is the first MFT equation~\eqref{eq:MFTq} for the dual system.
In order to get the second equation, we compute the time evolution of $\partial_k \pt$ using~\eqref{eq:RelPdual}, which yields
\begin{equation}
    \partial_t  \partial_k\pt
    = \partial_k \left[ 
    -\Dt(\qt) \partial_k^2 \pt
    - \frac{\st'(\qt)}{2} (\partial_k \pt)^2
    \right]
    \:.
\end{equation}
Integrating this relation from $-\infty$ to $k$, we get
\begin{equation}
    \partial_t  \pt =
    -\Dt(\qt) \partial_k^2 \pt
    - \frac{\st'(\qt)}{2} (\partial_k \pt)^2
    \:,
\end{equation}
which is the second equation~\eqref{eq:MFTp} for the dual system.

We have thus shown that the duality relation discussed in the main text at the level of the fluctuating hydrodynamics implies a duality of the MFT equations, with the transformation of $q$ and $p$ given by~(\ref{eq:MFTqDual},\ref{eq:RelPdual}).


%

\end{document}